\documentclass[preprint,review,12pt]{elsarticle}

\usepackage[a4paper, total={6in, 9.2in}]{geometry}
\usepackage[utf8]{inputenc}
\usepackage{mathabx}
\usepackage{graphicx}
\usepackage{siunitx}
\usepackage{hyperref}
\usepackage{adjustbox}
\usepackage{subcaption}
\usepackage{float}
\usepackage{placeins}
\usepackage{stfloats}
\usepackage{comment}

\journal{Ultramicroscopy}

\begin{document}
\begin{frontmatter}
\title{Correcting for probe wandering by precession path segmentation}

\author[add1]{Gregory Nordahl}
\ead{gregory.nordahl@ntnu.no}
\author[add2,add3]{Lewys Jones}
\author[add1]{Emil Frang Christiansen}
\author[add4]{Kasper Aas Hunnestad}
\author[add1]{Magnus Nord}

\address[add1]{Department of Physics, Norwegian University of Science and Technology (NTNU), 7491 Trondheim, Norway}
\address[add2]{School of Physics, Trinity College Dublin, Dublin 2, Ireland}
\address[add3]{Advanced Microscopy Laboratory, Centre for Research on Adaptive Nanostructures and Nanodevices (CRANN), Dublin 2, Ireland}
\address[add4]{Department of Materials Science and Engineering, Norwegian University of Science and Technology (NTNU), 7491 Trondheim, Norway}

\begin{abstract}
Precession electron diffraction has in the past few decades become a powerful technique for structure solving, strain analysis, and orientation mapping, to name a few. One of the benefits of precessing the electron beam, is increased reciprocal space resolution, albeit at a loss of spatial resolution due to an effect referred to as 'probe wandering'. Here, a new methodology of precession path segmentation is presented to counteract this effect and increase the resolution in reconstructed virtual images from scanning precession electron diffraction data. By utilizing fast pixelated electron detector technology, multiple frames are recorded for each azimuthal rotation of the beam, allowing for the probe wandering to be corrected in post-acquisition processing. Not only is there an apparent increase in the resolution of the reconstructed images, but probe wandering due to instrument misalignment is reduced, potentially easing an already difficult alignment procedure.
\end{abstract}


\end{frontmatter}

\section{Introduction}

Ever since the concept of precession electron diffraction (PED) was introduced by Vincent and Midgley almost 30 years ago \cite{Vincent1994}, the technique has seen an increase in popularity owing to the benefits of electron beam precession. Today it is an invaluable tool for structure determination \cite{Xie2008, White2010a, Klein2011}, strain measurements \cite{Rouviere2013}, and phase- and orientation mapping \cite{Rauch2010, Viladot2013}. Precessing the beam equates to the Ewald sphere sampling a volume of reciprocal space \cite{Eggeman2013}, exciting more reflections than conventional electron diffraction. At the same time, the individual reflection intensities appear kinematic-like \cite{White2010b, Eggeman2010}, with a monotonic intensity increase over a large thickness range with little variation between equally excited reflections \cite{Midgley2015}.


Although precession leads to an increased angular resolution in reciprocal space \cite{Koch2011}, experimental limitations mean that the spatial resolution suffers increasingly with higher precession angles; this is apparent in virtual bright field (VBF) reconstructions from scanning PED (SPED) data performed by Barnard et al. (2017) \cite{Barnard2017}. Due to the off-axis tilt of the beam, the probe is traversing the aberration surface in a circular pattern during precession, being displaced from the optical axis by a shift equal to the local gradient of the aberration function at each point on the precession azimuth \cite{Eggeman2013}. These shifts will lead to an increased time averaged and effective probe size. The probe shifts, often referred to as 'probe wandering' due to the periodic nature of the motion, are dominated by the unavoidable spherical aberration of the probe forming lens even under perfect alignment conditions \cite{Liao2012}. However, PED alignment is known to be difficult and a lot of research has gone into optimizing the procedure \cite{Eggeman2013, Barnard2017, Koch2011, Rouviere2013}. In a well aligned PED setup, the beam crossover and precession pivot point are both perfectly coincident on the sample surface \cite{Barnard2017}. A deviation of either would induce the same probe wandering as mentioned earlier, likely with a greater contribution than from spherical aberration. Although probe wandering is inherent to PED, the motion is periodic and, as will be shown, has the potential to be corrected.

In this work we present the methodology of precession path segmentation to counteract probe wandering, increasing the resolution in VBF images reconstructed from SPED scans. The basis of this methodology lies in fast pixelated detector technology, as the goal is to record several frames in quick succession for each azimuthal rotation of the beam. Results are presented from two applications of the methodology, one on a regular, well-aligned SPED scan, and one where a precession misalignment has been introduced to demonstrate the capabilities of the technique.

\section{Experimental methods}

To perform precession path segmentation, the first step is to align the instrument for SPED acquisition. These experiments were carried out on a JEOL JEM-2100F operating at \SI{200}{\kilo\volt}, equipped with a NanoMEGAS DigiSTAR precession scan generator and a Quantum Detectors single chip MerlinEM direct electron detector. The SPED alignment was done by following the double-rocking alignment outlined by Barnard et al. (2017) \cite{Barnard2017}, applied on a nanobeam diffraction \SI{0.5}{\nano\metre} probe with a \SI{1.2}{\milli\radian} convergence semi-angle. The precession angle was set to \SI{1.0}{\degree}, or approximately \SI{17.5}{\milli\radian}, and the precession frequency to \SI{100}{\hertz}. The selected precession angle is expected to bring the beam out of the aberration free area for such a non-corrected instrument.

After the alignment, scans were set up to record precession path segmentation datasets with $n=8$ segments, meaning that the detector had to record $8$ frames at each scan position. With the selected precession frequency, the dwell time was set to \SI{10}{\milli\second} to allow the beam one full rotation around the optical axis, and the scanner frame points were set to 256$\times$256. Consequently, the detector capture time was set to $10/\SI{8}{\milli\second} = \SI{1.25}{\milli\second}$, and frame captures to 2048$\times$256. In other words, the capture time was $1/n$ of the dwell time, and the number of X frames recorded was multiplied by $n$. To allow this increased amount of data to be recorded by the detector, the scanner flyback parameter had to be increased by a significant amount. The resulting dataset ends up being $n$ times larger than a regular SPED scan, and typically 2-3 times slower in recording, however during the flyback time the beam is off away to the side and adds minimal beam-damage to the region of interest. The number of segments was chosen so that diffraction patterns in each segment contained sufficient intensity for VBF reconstruction. To increase the amount of segments while maintaining adequate intensity levels in diffraction patterns, the precession frequency could be reduced.

Two SPED scans were recorded with our proposed precession path segmentation employed. One scan had a scan step size of \SI{5.6}{\nano\metre} and was performed with a proper precession alignment, while the other had a scan step size of \SI{10.2}{\nano\metre} and an intentional precession pivot point misalignment. The intentional misalignment was performed by adjusting X and Y scan coil amplitudes equally by \SI{0.1}{\percent} to offset the proper alignment. These intentionally lightly misaligned datasets serve as extreme tests of what might be a reasonable worst-case scenario with an experienced operator. The large 4D-STEM datasets were processed with the open-source Python libraries HyperSpy \cite{hyperspy} and pyxem \cite{pyxem}. Initially, the 2048$\times$256 dataset is sliced into 8 pieces, each of 256$\times$256 scan dimensions. From these slices, VBF reconstructions are created. Since probe wandering is equal at every scan point within a given precession angle, the features in the synthesized VBF images move relative to each other in a circular motion. This movement is corrected for by inserting the images as a stack into the SmartAlign plugin for Gatan Microscopy Suite, and performing rigid correction \cite{Jones2015} between the VBF images. Before rigid correction is performed, SmartAlign upscales the inserted images by a factor of two, which is possible due to the fact that the scan step sizes are much larger than the probe size. As a final step, the corrected images are summed to get a single 256$\times$256 VBF image output, and compared to a non-corrected VBF image sum. The non-corrected VBF image sum is equivalent to a conventional, non-segmented SPED scan VBF reconstruction.

\begin{figure}[!ht]
    \centering
    \includegraphics[width=\linewidth]{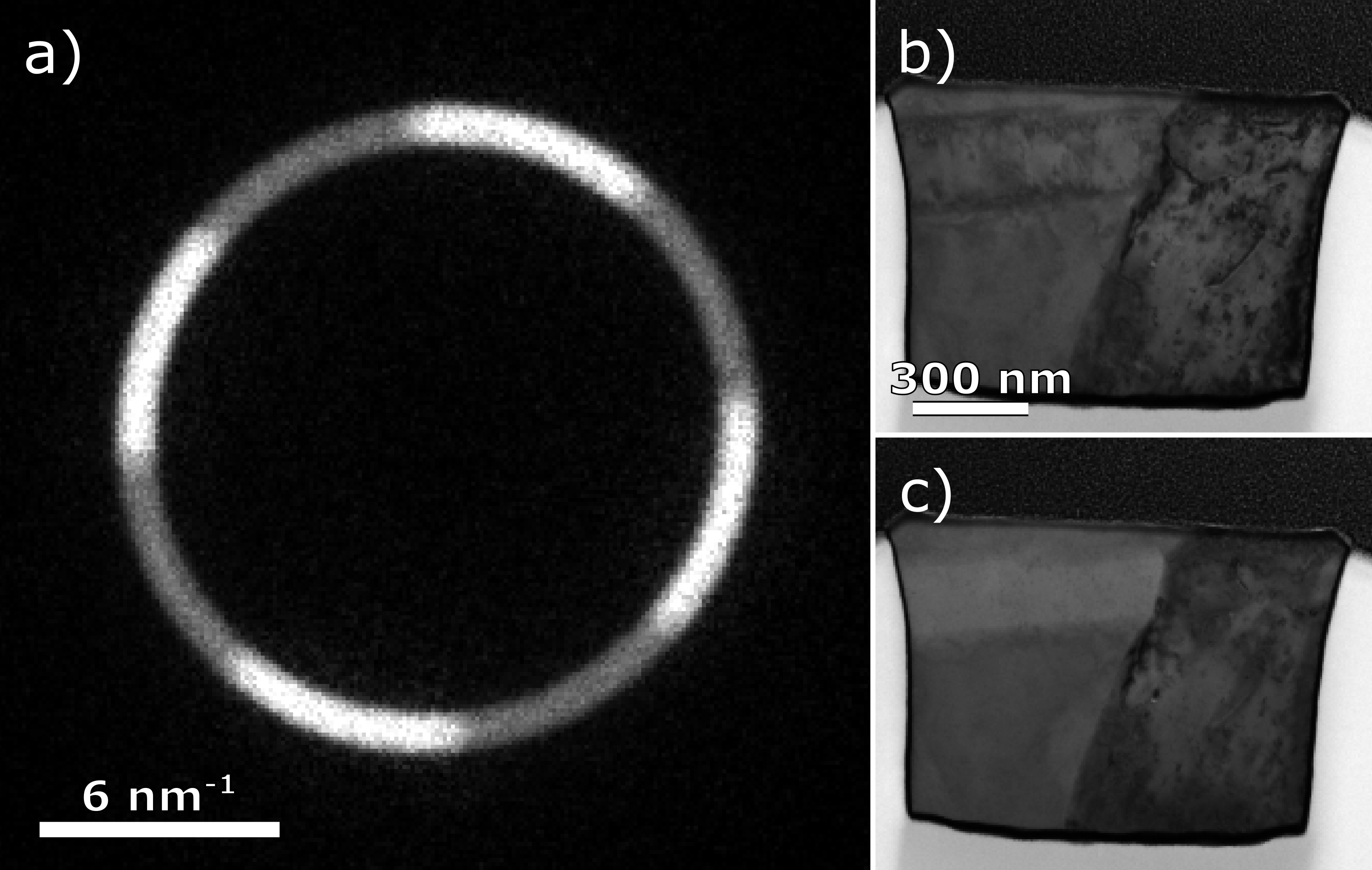}
    \caption{a) Eight precession path segments visualized in a diffraction pattern with de-rocking switched off. b) and c) VBF reconstructions corresponding to two different segments of the precession path.}
    \label{fig:segmentation}
\end{figure}

Figure \ref{fig:segmentation}a) is an overlap of $n=8$ individual diffraction patterns with precession de-rocking switched off, visualizing the segments of intensity integration. The methodology was tested on a focused ion beam cross-section lift-out of the integrated circuitry of a central processing unit (CPU) chip. Figures \ref{fig:segmentation}b) and \ref{fig:segmentation}c) are two of the eight VBF images reconstructed from the segmented scan with the smaller step size. The scan is of a crystalline region surrounded by amorphous areas, and inside the region there are large structural differences to be seen between the VBF images. This is due to the precession segments fulfilling different diffraction conditions on separate sides of the precession path.

\section{Results and discussion}

\begin{figure}[!ht]
    \centering
    \includegraphics[width=\linewidth]{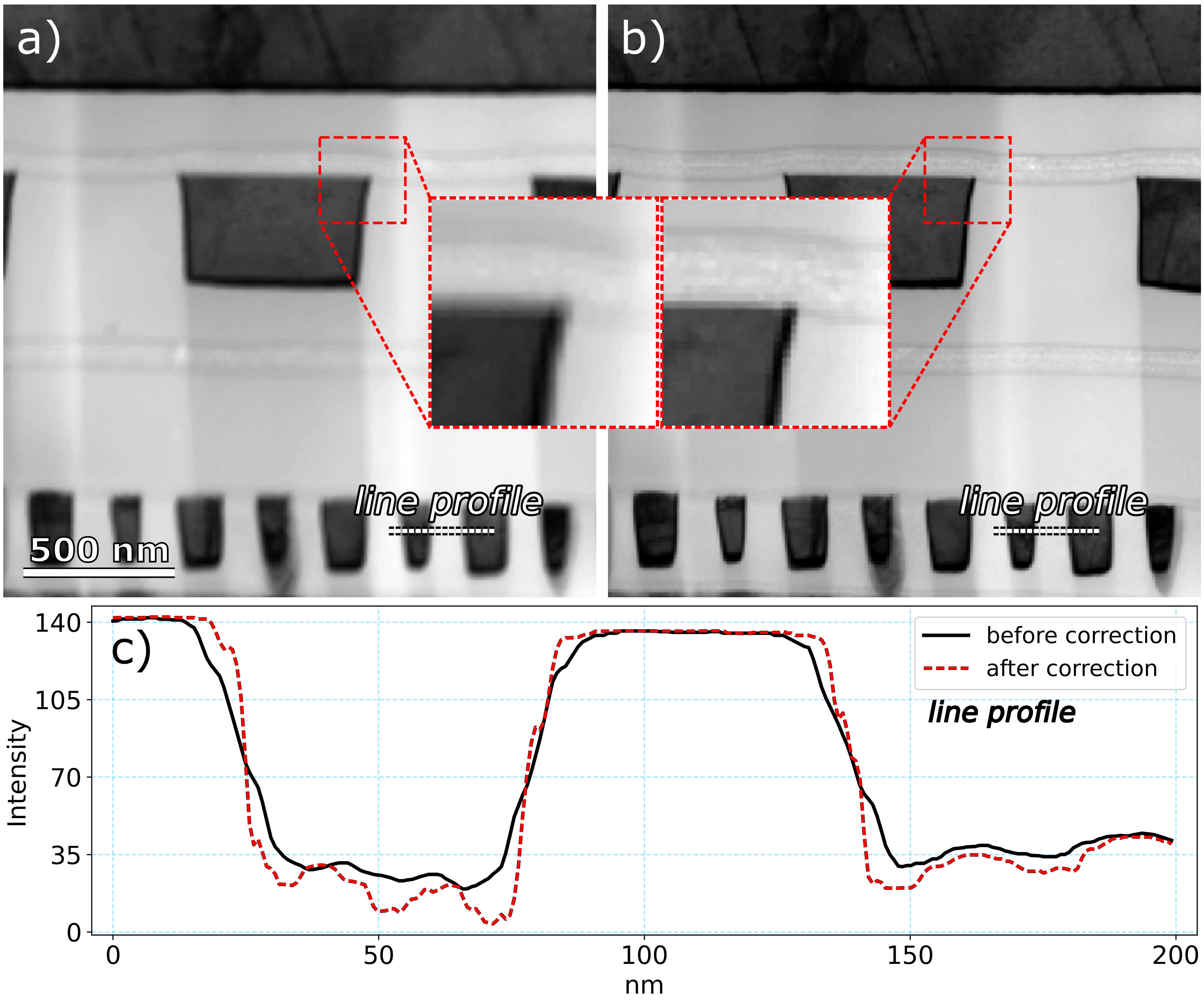}
    \caption{VBF images from a precession path segmented SPED scan where an intentional pivot point misalignment has been introduced, showing the sums a) before and b) after rigid correction. c) Line profiles from the same region in both scans are included.}
    \label{fig:results_pivot}
\end{figure}

The VBF images from the precession path segmented scan with a larger step size can be seen in Figure \ref{fig:results_pivot}, before and after rigid correction performed in SmartAlign. As previously mentioned, the scan had a small intentional pivot point misalignment introduced, and the probe shifts were estimated to be up to \SI{13}{\nano\metre} from the scan point center based on feature movement. The misalignment leads to blur in the VBF image when summed without correction, while the rigid corrected VBF image shows a noticeable contrast increase. The relative blurring is equivalent to the rigid corrected VBF image being convoluted with a Gaussian kernel 3 px in size with a standard deviation of $\sigma=7$. An intensity profile was extracted from equal positions in both images, covering a region with two crystalline features. The features show a larger dynamic range of intensity values and much higher structural variations in the rigidly corrected image as compared to the non-corrected one. The transition between a crystalline feature and the surrounding amorphous regions, i.e. the edges of the features, are also sharper. Least squares curve fitting was performed to estimate the increase in edge steepness between the uncorrected and corrected VBF images. An arctangent function, $A \cdot \arctan \left[ k \left( x-x_0 \right) \right]$ with fitting parameters $A, k, \text{and } x_0$, was fit on the rising edge of the feature as seen in the line profile between \SI{70}{\nano\metre} and \SI{90}{\nano\metre}. Results of the curve fittings had slope parameters, $k$, of $0.15 \pm 0.01$ for the non-corrected VBF edge, and $0.38 \pm 0.03$ for the corrected one, in other words a 2.5 times increase in edge steepness.

\begin{figure}[!ht]
    \centering
    \includegraphics[width=\linewidth]{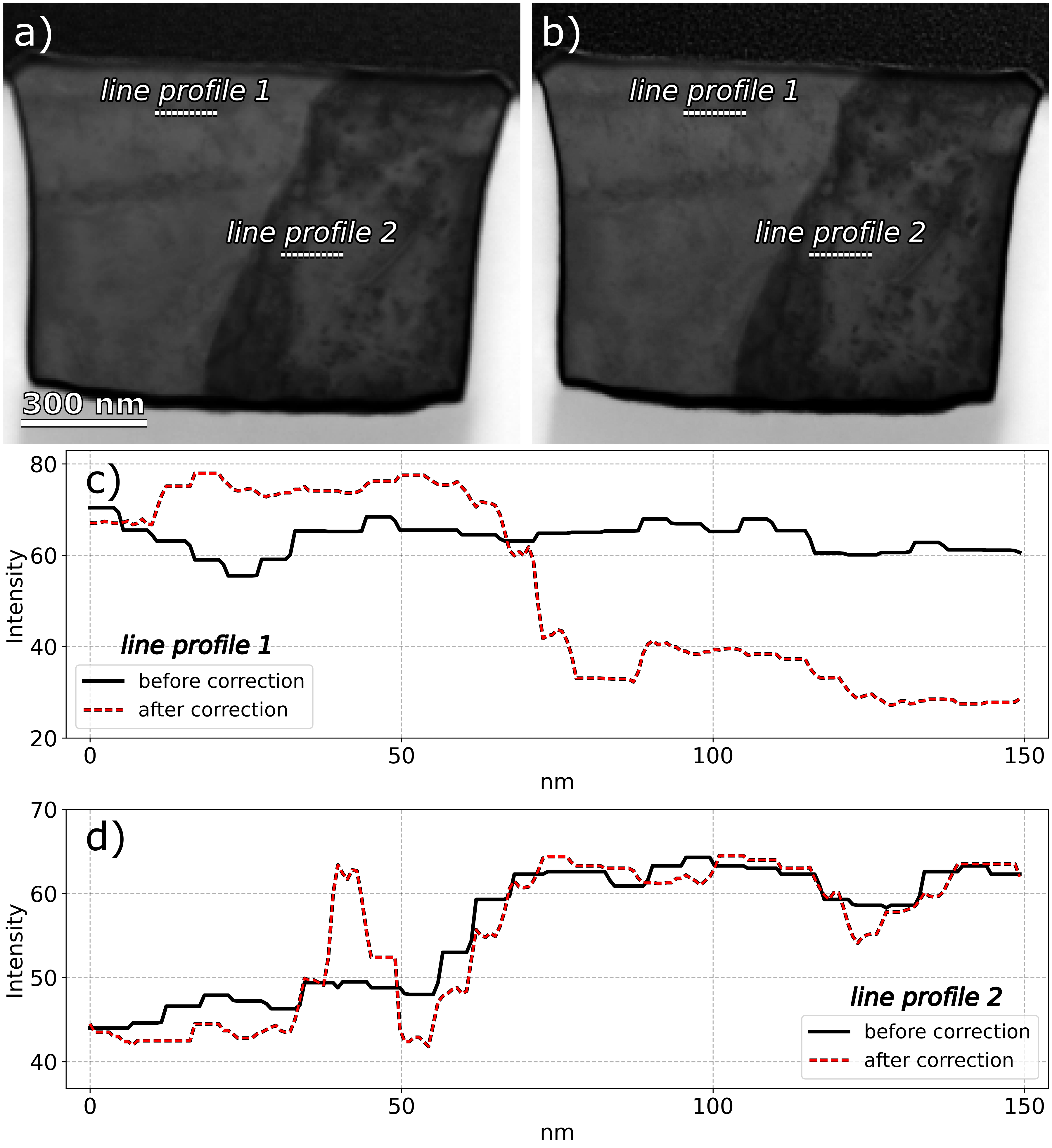}
    \caption{VBF images from a precession path segmented SPED scan with a proper alignment. Sum of individual VBF a) before and b) after rigid correction. c) and d) Intensity profiles from two different regions have been extracted from the two VBF sums.}
    \label{fig:results_small}
\end{figure}

A comparison of VBF images before and after rigid correction, from the precession path segmented scan with a smaller step size, can be seen in Figure \ref{fig:results_small}. As the scan did not have an intentional pivot point misalignment introduced as the previous one, visually there is a small, but noticeable, difference between the before- and after corrected images. Here the probe shifts have been estimated to be up to \SI{7}{\nano\metre}. Two intensity profiles have been extracted from equal positions in both images, as seen in Figure \ref{fig:results_small} b) and c). The intensity range in these profiles are on a much smaller scale than the profile in Figure \ref{fig:results_pivot}c), allowing us to see the small structural variations inside the crystalline region. The first profile shows a monotone intensity over the whole profile length in the non-corrected image, while the corrected image appears to have a broad intensity extension with a large intensity dip centered around the \SI{70}{\nano\metre} point, suggesting a feature edge. In the second profile, the corrected image shows larger structural variations on a smaller scale than the previous profile, such as the intensity spike around the \SI{40}{\nano\metre} point. Both profiles hint at increased contrast after rigid correction, while the non-corrected VBF image appears more blurred from the probe wandering.

The experimental results above show that the resolution can be increased in VBF images when probe wandering of \SIrange{7}{13}{\nano\metre} is present. Although the scan that showed the smallest wandering was aligned to the best extent of the operator, it is difficult to state to what degree the spherical aberration of the probe forming lens influences the shifts. It has been reported that shifts down to \SI{1}{\nano\metre} are possible for an aberration corrected instrument \cite{Eggeman2013}, which would imply that the \SI{7}{\nano\metre} shifts observed in the non-corrected instrument are to a large degree influenced by the spherical aberration. The observed probe wandering does not reflect the true nature of the shifts present in a SPED setup, as there are likely shifts within the individual segments of intensity integration that were averaged out. To achieve the most precise results, the precession path should be segmented until each segment resembles a single probe. For a probe with semi-convergence angle, $\alpha = \SI{1.2}{\milli\radian}$ and with a precession angle $\phi = \SI{17.5}{\milli\radian}$, to achieve single-probe segments one would have to segment the scan $2\pi\phi/2\alpha \approx 46$ times. This would drastically increase the scan size, and each individual segment's diffraction patterns would have very low intensity unless compensated for by having a much smaller precession frequency than the \SI{100}{\hertz} utilized. The final result would imitate a $\phi$-tilt series, albeit being recorded instantaneously with all information present in one single scan. It is worth pointing out that although precession path segmentation increases the apparent resolution in VBF images, it does so by counteracting the effect of probe wandering from the beam precessing on an aberration surface. The optimum case, as described above, will be achieved with a segmented scan where each segment of intensity integration takes the appearance of a single beam, but the resolution will not be increased further than what is achievable without precession.


The methodology in this article has been applied solely to VBF reconstructions, since probe wandering is straightforward recognizable as feature movement between the images corresponding to different segments of the precession path. However, it is also possible to correct for probe wandering by overlapping diffraction patterns from different segments that correspond to the same spatial position of the probe. Since probe wandering is present, the individual segments of the scan need to be shifted by a fixed amount in both spatial directions. The shifts can be estimated based on first performing VBF reconstructions and measuring the displacement of recognizable features between the different VBF images, as done previously. More optimally one could calculate the shifts based on experimental parameters and the aberrations of the probe forming lens \cite{Eggeman2013}, and attempt to match probe step size in the scan for accurate fitting. This is a challenge as one would essentially attempt to reconstruct a circular path using a finite amount of pixels, which would work for horizontal and vertical pixels, albeit not diagonal elements. To better fit the distance between pixels on a diagonal, one could set up a scan with a step size a fraction of the expected probe shifts.
\section{Conclusion}

To conclude, we have presented a methodology to increase the resolution in VBF images from SPED data by reducing the effect of probe wandering, a by-product of electron beam precession which causes image blur. By segmenting the precession path, and performing VBF reconstructions on individual segments, apparent feature movement between images stemming from probe shifts is rigidly corrected in SmartAlign. The final output, which is a sum of the rigidly corrected VBF images, shows a significant reduction in blur, particularly evident in the small structural elements that appeared after correction. The methodology has proven effective in correcting the shifts due to instrumental misalignment, which might help to ease the alignment procedure for the operator, or alternatively, allow for more throughput on a microscopy session as the alignment requirements are less strict for high quality imaging.



\section*{Declaration of Competing Interest}

The authors declare no competing interest.

\section*{Data availability}

The repository for raw datasets, SmartAlign corrected images, and scripts to process both of these can be found at \url{https://doi.org/10.5281/zenodo.7319130}.

\section*{Acknowledgements}

We wish to acknowledge the support from the Research Council of Norway for the Norwegian Center for Transmission Electron Microscopy, NORTEM (197405), the Norwegian Micro- and Nano-Fabrication Facility, NorFab (295864), and In-situ Correlated Nanoscale Imaging of Magnetic Fields in Functional Materials, InCoMa (315475). LJ acknowledges funding from Science Foundation Ireland grant URF/RI/191637.

\bibliographystyle{elsarticle-num}
\bibliography{references.bib}

\begin{thebibliography}{10}
\expandafter\ifx\csname url\endcsname\relax
  \def\url#1{\texttt{#1}}\fi
\expandafter\ifx\csname urlprefix\endcsname\relax\def\urlprefix{URL }\fi
\expandafter\ifx\csname href\endcsname\relax
  \def\href#1#2{#2} \def\path#1{#1}\fi

\bibitem{Vincent1994}
R.~Vincent, P.~Midgley, Double conical beam-rocking system for measurement of
  integrated electron diffraction intensities, Ultramicroscopy 53~(3) (1994)
  271--282.
\newblock \href {https://doi.org/https://doi.org/10.1016/0304-3991(94)90039-6}
  {\path{doi:https://doi.org/10.1016/0304-3991(94)90039-6}}.

\bibitem{Xie2008}
D.~Xie, C.~Baerlocher, L.~B. McCusker, {Combining precession electron
  diffraction data with X-ray powder diffraction data to facilitate structure
  solution}, Journal of Applied Crystallography 41~(6) (2008) 1115--1121.
\newblock \href {https://doi.org/https://doi.org/10.1107/S0021889808034377}
  {\path{doi:https://doi.org/10.1107/S0021889808034377}}.

\bibitem{White2010a}
T.~A. White, M.~S. Moreno, P.~A. Midgley, Structure determination of the
  intermediate tin oxide sn3o4 by precession electron diffraction, Zeitschrift
  für Kristallographie 225~(2-3) (2010) 56--66.
\newblock \href {https://doi.org/https://doi.org/10.1524/zkri.2010.1210}
  {\path{doi:https://doi.org/10.1524/zkri.2010.1210}}.

\bibitem{Klein2011}
H.~Klein, {Precession electron diffraction of Mn${\sb 2}$O${\sb 3}$ and
  PbMnO${\sb 2.75}$: solving structures where X-rays fail}, Acta
  Crystallographica Section A 67~(3) (2011) 303--309.
\newblock \href {https://doi.org/https://doi.org/10.1107/S0108767311009512}
  {\path{doi:https://doi.org/10.1107/S0108767311009512}}.

\bibitem{Rouviere2013}
J.-L. Rouviere, A.~Béché, Y.~Martin, T.~Denneulin, D.~Cooper, Improved strain
  precision with high spatial resolution using nanobeam precession electron
  diffraction, Applied Physics Letters 103~(24) (2013) 241913.
\newblock \href {https://doi.org/https://doi.org/10.1063/1.4829154}
  {\path{doi:https://doi.org/10.1063/1.4829154}}.

\bibitem{Rauch2010}
E.~F. Rauch, J.~Portillo, S.~Nicolopoulos, D.~Bultreys, S.~Rouvimov, P.~Moeck,
  Automated nanocrystal orientation and phase mapping in the transmission
  electron microscope on the basis of precession electron diffraction,
  Zeitschrift für Kristallographie 225~(2-3) (2010) 103--109.
\newblock \href {https://doi.org/https://doi.org/10.1524/zkri.2010.1205}
  {\path{doi:https://doi.org/10.1524/zkri.2010.1205}}.

\bibitem{Viladot2013}
D.~Viladot, M.~Véron, M.~Gemmi, F.~Peiró, J.~Portillo, S.~Estradé,
  J.~Mendoza, N.~Llorca-Isern, S.~Nicolopoulos, Orientation and phase mapping
  in the transmission electron microscope using precession-assisted diffraction
  spot recognition: state-of-the-art results, Journal of Microscopy 252~(1)
  (2013) 23--34.
\newblock \href {https://doi.org/https://doi.org/10.1111/jmi.12065}
  {\path{doi:https://doi.org/10.1111/jmi.12065}}.

\bibitem{Eggeman2013}
A.~S. Eggeman, J.~S. Barnard, P.~A. Midgley, Aberration-corrected and
  energy-filtered precession electron diffraction, Z. Kristallogr. Cryst. Mater
  228~(1) (2013) 43--50.
\newblock \href {https://doi.org/https://doi.org/10.1524/zkri.2013.1565}
  {\path{doi:https://doi.org/10.1524/zkri.2013.1565}}.

\bibitem{White2010b}
T.~A. White, A.~S. Eggeman, P.~A. Midgley, Is precession electron diffraction
  kinematical? part i:: "phase-scrambling" multislice simulations,
  Ultramicroscopy 110~(7) (2010) 763--770.
\newblock \href
  {https://doi.org/https://doi.org/10.1016/j.ultramic.2009.10.013}
  {\path{doi:https://doi.org/10.1016/j.ultramic.2009.10.013}}.

\bibitem{Eggeman2010}
A.~Eggeman, T.~White, P.~Midgley, Is precession electron diffraction
  kinematical? part ii: A practical method to determine the optimum precession
  angle, Ultramicroscopy 110~(7) (2010) 771--777.
\newblock \href
  {https://doi.org/https://doi.org/10.1016/j.ultramic.2009.10.012}
  {\path{doi:https://doi.org/10.1016/j.ultramic.2009.10.012}}.

\bibitem{Midgley2015}
P.~A. Midgley, A.~S. Eggeman, {Precession electron diffraction {--} a topical
  review}, IUCrJ 2~(1) (2015) 126--136.
\newblock \href {https://doi.org/https://doi.org/10.1107/S2052252514022283}
  {\path{doi:https://doi.org/10.1107/S2052252514022283}}.

\bibitem{Koch2011}
C.~T. Koch, Aberration-compensated large-angle rocking-beam electron
  diffraction, Ultramicroscopy 111~(7) (2011) 828--840, special Issue: J.
  Spence's 65th birthday.
\newblock \href
  {https://doi.org/https://doi.org/10.1016/j.ultramic.2010.12.014}
  {\path{doi:https://doi.org/10.1016/j.ultramic.2010.12.014}}.

\bibitem{Barnard2017}
J.~S. Barnard, D.~N. Johnstone, P.~A. Midgley, High-resolution scanning
  precession electron diffraction: Alignment and spatial resolution,
  Ultramicroscopy 174 (2017) 79--88.
\newblock \href
  {https://doi.org/https://doi.org/10.1016/j.ultramic.2016.12.018}
  {\path{doi:https://doi.org/10.1016/j.ultramic.2016.12.018}}.

\bibitem{Liao2012}
Y.~Liao, L.~D. Marks, On the alignment for precession electron diffraction,
  Ultramicroscopy 117 (2012) 1--6.
\newblock \href
  {https://doi.org/https://doi.org/10.1016/j.ultramic.2012.03.021}
  {\path{doi:https://doi.org/10.1016/j.ultramic.2012.03.021}}.

\bibitem{hyperspy}
F.~de~la Peña, E.~Prestat, V.~T. Fauske, P.~Burdet, T.~Furnival,
  P.~Jokubauskas, J.~Lähnemann, M.~Nord, T.~Ostasevicius, K.~E. MacArthur,
  et~al., hyperspy/hyperspy: Release v1.6.4 (07 2021).
\newblock \href {https://doi.org/https://doi.org/10.5281/zenodo.5082777}
  {\path{doi:https://doi.org/10.5281/zenodo.5082777}}.

\bibitem{pyxem}
D.~N. Johnstone, P.~Crout, M.~Nord, J.~Laulainen, S.~Høgås, E.~Opheim,
  B.~Martineau, C.~Francis, T.~Bergh, E.~e.~a. Prestat, pyxem/pyxem: pyxem
  0.13.3 (07 2021).
\newblock \href {https://doi.org/https://doi.org/10.5281/zenodo.5075520}
  {\path{doi:https://doi.org/10.5281/zenodo.5075520}}.

\bibitem{Jones2015}
L.~Jones, H.~Yang, T.~J. Pennycook, M.~S.~J. Marshall, S.~Van~Aert, N.~D.
  Browning, M.~R. Castell, P.~D. Nellist, Smart align—a new tool for robust
  non-rigid registration of scanning microscope data, Advanced Structural and
  Chemical Imaging 1 (2015) 8--23.
\newblock \href {https://doi.org/https://doi.org/10.1186/s40679-015-0008-4}
  {\path{doi:https://doi.org/10.1186/s40679-015-0008-4}}.

\end{thebibliography}

\end{document}